\RequirePackage{lineno}
\documentclass[aps, prl,
preprint,
floatfix,
superscriptaddress,showpacs]{revtex4-1}
\usepackage{t1enc}
\usepackage{hyperref}
\usepackage[english]{babel}
\usepackage{graphicx}
\usepackage{xspace}
\usepackage{natbib}

\usepackage[usenames,dvipsnames]{color}

\providecommand{\smhb}{samarium hexaboride\xspace}
\providecommand{\smbsix}{\ensuremath{\mathrm{SmB}_6}\xspace}
\providecommand{\Xbar}{\ensuremath{\bar{X}}\xspace}
\providecommand{\Mbar}{\ensuremath{\bar{M}}\xspace}
\providecommand{\X}{\ensuremath{X}\xspace}
\providecommand{\Gbar}{\ensuremath{\bar{\Gamma}}\xspace}

\providecommand{\etal}{\emph{et al.}\xspace}
\providecommand{\ie}{i.e.,\xspace}

\providecommand{\sHfivetwo}{\ensuremath{^6H_{5/2}}\xspace}
\providecommand{\sHseventwo}{\ensuremath{^6H_{7/2}}\xspace}
\providecommand{\EF}{\ensuremath{E_\mathrm{F}}\xspace}

\providecommand{\twodim}{two-dimensional\xspace}
\providecommand{\threedim}{three-dimensional\xspace}
\providecommand{\threed}{3-D\xspace}

\providecommand{\kperp}{\ensuremath{k_{\perp}}\xspace}

\newcommand{\oli}[1]{{\color{black} #1}}
\newcommand{\oliv}[1]{{\color{black} #1}}
\newcommand{\emile}[1]{{\color{black} #1}}

\begin{document}
\title{Nodeless Hybridization as Proof of Trivial Topology in Samarium Hexaboride}

\def\SuppNoteDim{\oli{1}}
\def\SuppFigDim{\oli{S1}}
\def\SuppNoteRes{\oli{2}}
\def\SuppFigRes{\oli{S2}}

\author{E. D. L. Rienks}
\email{emile.rienks@helmholtz-berlin.de}
\author{P. Hlawenka}
\affiliation{Helmholtz-Zentrum Berlin f\"ur Materialien und Energie, Albert-Einstein-Str.~15, 12489 Berlin, Germany}
\author{J. S\'anchez-Barriga}
\affiliation{Helmholtz-Zentrum Berlin f\"ur Materialien und Energie, Albert-Einstein-Str.~15, 12489 Berlin, Germany}
\affiliation{IMDEA Nanociencia, c/ Faraday 9, Campus de Cantoblanco, 28049 Madrid, Spain}
\author{E. Schierle}
\affiliation{Helmholtz-Zentrum Berlin f\"ur Materialien und Energie, Albert-Einstein-Str.~15, 12489 Berlin, Germany}
\author{M. Jugovac}
\affiliation{Elettra - Sincrotrone Trieste, S.S. 14 - km 163.5, Basovizza, 34149 Trieste, Italy}
\author{P. Perna}
\author{I. Cojocariu}
\affiliation{IMDEA Nanociencia, c/ Faraday 9, Campus de Cantoblanco, 28049 Madrid, Spain}
\author{Kai Chen}
\altaffiliation[Present address: ] {National Synchrotron Radiation Laboratory, University of Science and
Technology of China, Hefei 230026, China}
\author{K. Siemensmeyer}
\author{E. Weschke}
\author{A. Varykhalov}
\affiliation{Helmholtz-Zentrum Berlin f\"ur Materialien und Energie, Albert-Einstein-Str.~15, 12489 Berlin, Germany}
\author{N. Y. Shitsevalova}
\author{V. B. Filipov}
\affiliation{Institute for Problems of Materials Science, National Academy of Sciences of Ukraine, Kiev, 03142 Ukraine}
\author{S. Gab\'ani}
\author{K. Flachbart}
\affiliation{Institute of Experimental Physics, Slovak Academy of Sciences, Košice, 04001 Slovakia}
\author{O. Rader}
\email{rader@helmholtz-berlin.de}
\affiliation{Helmholtz-Zentrum Berlin f\"ur Materialien und Energie, Albert-Einstein-Str.~15, 12489 Berlin, Germany}

\date{\today}

\begin{abstract}
Calculations unanimously predict \smhb to be a topological Kondo insulator\emile{,} the first topological insulator driven by strong electron correlation. As of today it appears also experimentally established as the only representative of \emile{this} material \emile{class}. Here, we investigate the three-dimensional band structure of \smbsix and show that it is incompatible with a topological Kondo insulator which must have hybridization nodes at high-symmetry points. In addition we clarify the  remaining questions concerning \emile{the nature of the} surface states with new data. We address consequences for the search for correlated topological insulators. 
\end{abstract}

\maketitle

\section{Introduction}
In the 21st century, solid-state physics has been shaped primarily by topology and strong electron correlations—two largely distinct fields. Topological band insulators were first predicted theoretically and later confirmed experimentally \cite{Hasan:2010ku}, with an ever-growing number leading to the realization that over 27\%\ of all natural materials are topological \cite{Vergniory:2019aa}. Topological band theory relies on single-particle Hamiltonians to predict topological invariants \cite{BansilRMP16}, and angle-resolved photoelectron spectroscopy (ARPES) \oli{\cite{ZhangARPES22}} has become the leading technique for their investigation \cite{Sanchez-Barriga:2024aa}. 

Topological insulators driven by strong electron correlation could overcome key limitations of conventional topological insulators, particularly their small inverted bulk band gaps, which restrict their use in semiconductor electronics. Coulomb interactions can significantly enhance these gaps, centering them also around the Fermi level without requiring specific stoichiometry \oli{and control of dopants} \cite{Dzero:2013cd}. Among strongly correlated topological insulators \smbsix remains the most extensively studied and the only widely accepted candidate.  
While theoretical predictions unambiguously identify \smbsix as a topologically nontrivial material, recent survey articles also strongly affirm its experimental confirmation as a topological insulator \cite{Rachel:2018un, Li:2020jh, Checkelsky:2024aa}.

\smbsix has been predicted as a topological Kondo insulator in a series of publications \cite{Dzero10, Alexandrov13, Takimoto11, LuF13}. In a Kondo insulator, the band gap arises from the hybridization of itinerant conduction electrons with localized $f$-electrons, leading to the elegant and pioneering idea \oliv{by Dzero \etal} of a band inversion between even-parity itinerant $d$-states and odd-parity localized $f$-states, \oliv{the topological Kondo insulator} \cite{Dzero10}. Band structure calculations \oli{specifically} predict \smbsix as a strong topological insulator with an odd number of Dirac cone surface states \cite{Alexandrov13, Takimoto11, LuF13}. Historically, \smbsix\ \oliv{had been} identified as the first mixed-valent compound and the first Kondo insulator. Its well-documented conduction anomaly — where the resistance attains a plateau below 4 K — was later attributed to surface conduction \cite{Wolgast:2013ih, KimDJSmGdB6}, consistent with the expected behavior of a topological insulator.

\oliv{After the} prediction by Dzero \etal , surface states were reported \oliv{in ARPES} at both the center (\Gbar) \cite{XuPRB13,JiangNC13} and edge (\Xbar) of the (100) surface Brillouin zone. The \Xbar state manifests as a prominent elliptical Fermi surface contour \cite{XuPRB13,JiangNC13,Neupane13,Frantzeskakis13,Denlinger13126637,Min14}. This was taken as proof of the nontrivial topology of \smbsix by most \cite{XuPRB13,JiangNC13,Neupane13,Denlinger13126637,Min14}, even though it was pointed out that the situation is not settled and a Dirac cone not observed  \cite{Frantzeskakis13}.
Xu \etal subsequently reported a spin texture consistent with a topological surface state at \Xbar \cite{XuNC14}.

\oliv{Subsequently,}  we proposed an entirely trivial explanation for the surface states. The innermost \Gbar state shows a parabolic dispersion which develops a Rashba splitting for boron termination and disappears with adsorbates. The \Xbar state is the $5d$-$4f$ hybridized dispersion but confined to 2D like in semiconductor band bending, however, on an energy scale of 10 meV \cite{Hlawenka:2018fj}. Our trivial explanation of the surface states is fully consistent with other ARPES and scanning tunneling microscopy/spectroscopy (STM/STS) studies \cite{Herrmann:2020ju}. Furthermore, our trivial explanation of the surface states remains largely unchallenged \cite{Li:2020jh}, while the conclusion that ARPES and spin-resolved ARPES confirmed the nontrivial topology of \smbsix was based on arguments that we will critically examine below. 

\oliv{ Miao \etal \cite{Miao21}  reported that the topological surface states of \smbsix   remain robust in ARPES even with 30\%\ Ce   and 20\%\ Eu substitution, the latter inducing antiferromagnetism. Notably, surface conductivity also persists with magnetic Eu doping \cite{Anisimov24}.} 

To date most experimental studies have attempted to address the topological nature of \smbsix through the properties of its surface states and concluded on a nontrivial topology \cite{XuPRB13, JiangNC13, Neupane13, Denlinger13126637, Min14, XuNC14, Ohtsubo:2019hk, Pirie:2019gr, Miao21, Aishwarya22,Pirie:2023aa}.
In this article we will take a more fundamental approach by focusing on the \threedim band structure \oliv{since it} \oli{determines the topology}. We will show that the \threed electronic structure of \smbsix violates the fundamental property of a topological Kondo insulator that hybridization must vanish at high-symmetry points.

\section{Results}

\subsection{\threed Electronic Structure}

Hybridization between localized and itinerant states of opposite parity is the crucial ingredient of a topological Kondo insulator: It gives rise to an inversion-symmetric insulator with a mixed parity occupied band structure \cite{Dzero10,Dzero:2016ky}. An important consequence of the opposite parity of the involved states is that there is no mixing at high symmetry points {\oliv{\cite{Dzero10, Takimoto11,  Alexandrov13, LuF13, Dzero:2016ky, Takimoto16}}}.
\emile{This follows from the fact that hybridization, which is periodic in the reciprocal lattice vector $\mathbf{G}$, must be odd under the parity transformation to mix opposite parity states \cite{Dzero:2016ky}.}

\begin{figure}
    \centering
    \includegraphics[width=.9\linewidth]{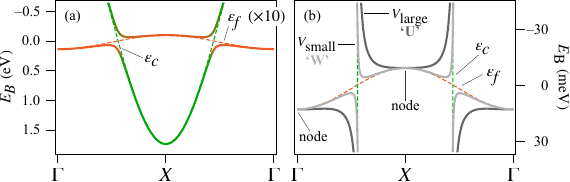}
    \caption{(a) Minimal representation of the topological Kondo insulator in which hybrid bands (solid curves) are formed by mixing opposite-parity localized ($\varepsilon_f$, dashed orange) and itinerant ($\varepsilon_c$, dashed green curve) states \cite{Dzero:2016ky}. Localized band width increased by factor 10 for clarity. (b) Hybrid states obtained for weaker ($V_\mathrm{small}$ solid light gray) and stronger mixing ($V_\mathrm{large}$ dark gray curves).
    \oli{The relative hybridization strength \oliv{(at $T=0$ K)} varies in the calculations, ranging from small (W-shape) \cite{LuF13,Denlinger13126637} to  large (U-shape) \cite{Iraola24}. }
    \oli{Essential is that} opposite parity implies hybridization nodes at high-symmetry points where hybrid states are pinned to their unmixed constituents.}
    \label{fig:schematic}
\end{figure}

This implies that the localized state must have an at least weak band dispersion with a phase opposite to that of the itinerant band, otherwise hybridization cannot give rise to a band gap (see Fig.~\ref{fig:schematic}). Another way of phrasing this requirement is that the hybridization gap in a topological Kondo insulator opens by mixing at a location in momentum space \emph{between} high symmetry points, while the dispersion \emph{at} those points must remain unaffected. In the following we will assess whether the \threedim band structure of \smbsix complies with this condition.

What part of the \threed band structure can be studied with ARPES? Min \etal and Denlinger \etal reported on the appearance of the bulk conduction band around the \X point at elevated temperature \cite{Min14,Denlinger13126637}.  Since the conduction band does not overlap with photoemission intensity from other states, \oli{it is straightforward to analyze}. In the aforementioned studies it was assumed, but not explicitly shown, that this feature is really \threedim . Photon energy dependent measurements in Fig. \SuppFigDim\ show that the conduction band feature is observed only around bulk \X points, thereby confirming its three-dimensional nature.

\subsubsection{Temperature Dependence}

Having established the \threed nature of the conduction band feature, we will now focus on its temperature dependent behavior in the temperature range where \smbsix becomes bulk insulating by $d$--$f$ hybridization.

\begin{figure}
    \centering
    \includegraphics[width=1.0\linewidth]{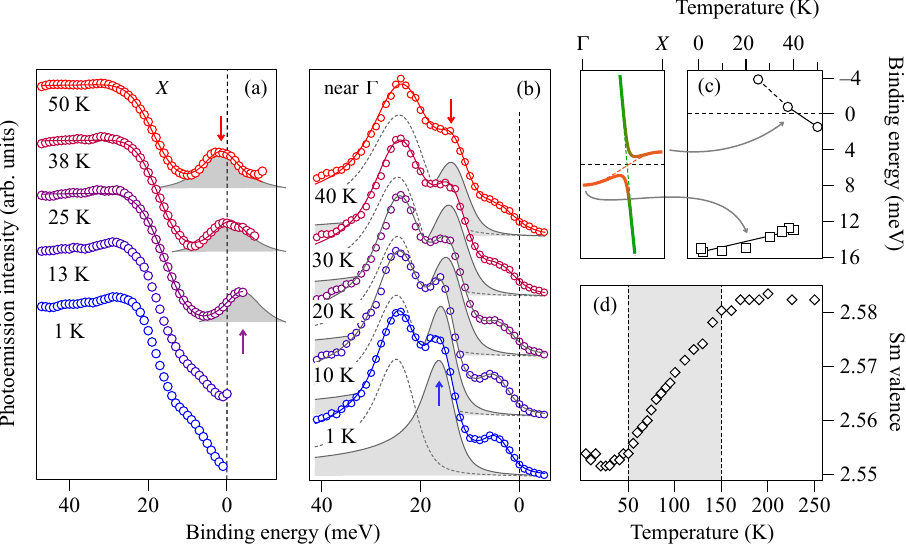}
    \caption{Temperature dependence of the \threed and bulk-like features at \Xbar and \Gbar.
    (a) Temperature dependent photoemission spectra at \Xbar divided by the Fermi-Dirac function. Solid line is a fit to the spectrum using a Lorentzian component (filled gray curves) for the conduction band.
    (b) Corresponding spectra at \Gbar fitted with a superposition of Doniach-Sunjic functions. Surface (dashed) and more bulk-like Sm $4f$ components (filled gray curves) are indicated.
    (c) Positions of the peaks highlighted in (a) and (b) versus temperature.
    (d) Temperature-dependent Sm valence determined with M$_5$ edge X-ray absorption spectroscopy.}
    \label{fig:bulktd}
\end{figure}

    Our experimental results in Fig.~\ref{fig:bulktd}(a) are in agreement with the earlier studies by Min \etal and Denlinger \etal : The conduction band's photoemission maximum at \Xbar has a positive binding energy above $T=50$~K. It shifts above \EF and becomes entirely depopulated upon cooling.
%
    In the temperature range where a sufficient fraction of the state is thermally populated (from 50 to 25 K), we can observe a reduction in binding energy by approximately 5 meV. Given its finite width, the conduction band is likely shifting by another similar amount upon further cooling to 1 K. We can thus estimate the shift from 50 to 1 K to be approximately 10 meV.
  %
    Accompanying the shift to lower binding energy of the conduction band feature, we observe the \emph{opposite} behavior for the bulk-like states at \Gbar (Fig.~\ref{fig:bulktd}). We have found the lower-binding energy peak of the \oli{two peaks} at \Gbar to be more bulk-like by means of adsorption experiments, see Hlawenka \etal \cite{Hlawenka:2018fj}.
    %
    We thus observe a non-rigid shift of the band structure upon cooling that displaces spectral weight away from \EF . Since transport measurements show a strong reduction in conductivity in this temperature window, this change apparently causes the material to become insulating. We note that this is different from what was suggested in the earlier study by Denlinger \etal \cite{Denlinger13126637}, who claim that the conduction and valence bands shift to lower binding energy with decreasing temperature in a parallel fashion.
   
    Our results reveal a non-rigid band shift that appears compatible with the temperature dependent Sm valence determined with $L_3$ edge X-ray absorption spectroscopy by Mizumaki \etal \cite{Mizumaki:2009ju}. We have obtained a very similar result for absorption near the $M_5$ edge ($\sim$1.08 keV) given in Fig.~\ref{fig:bulktd}(d): The valence remains fairly constant upon cooling from room temperature to 150 K and then decreases upon further cooling to 50 K. In the lowest temperature window between 50 and 2 K, in which \smbsix becomes bulk insulating, the valence \oliv{is rather constant and may} show a slight increase.   The relatively weak dependence of the valence \oliv{below 50 K} fits well with the non-rigid band change of the $f$-like sections of the quasi-particle bands we observe with ARPES.


\subsubsection{Connection Photoemission --- \threed Band Structure}

\begin{figure}
    \centering
    \includegraphics[width=0.9\linewidth]{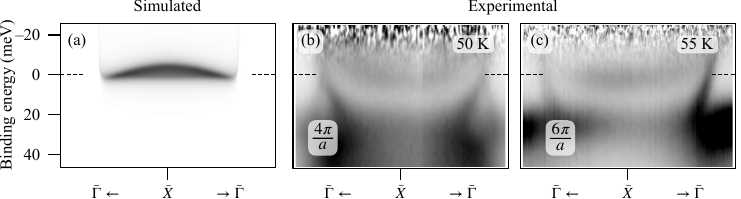}
    \caption{(left) Simulated photoemission intensity with the plano-convex shape required for a  hybridization induced shift that would be compatible with a topological Kondo insulator under the additional assumption of strong \kperp broadening. (right) Experimental photoemission intensity revealing a U-shaped dispersion for \kperp $=4.0\pi/a$ as well as $6.0 \pi/a$.}
    \label{fig:cbshape}
\end{figure}

    Is the observed shift at \Xbar caused by a change of the \threed band structure at \X , or can it be a result of changes near but not persisting at \X ?
    We recall that in the scenario of the topological Kondo insulator the conduction band dispersion must be pinned at \X regardless of hybridization strength $V$.
    
  \oli{This is not so in the topologically trivial standard model of Kondo insulators where the dispersive conduction band hybridizes with a flat $f$ band \cite{Riseborough00, Dzero:2016ky}. The hybridization matrix element $V$  induces a  large direct gap and a small \oliv{indirect gap   between the zone center and the zone boundary}. In the picture of the Kondo effect, strong electron scattering off the local moments causes below the Kondo temperature the formation of spin singlets. As the temperature increases, $V$ is suppressed due to spin disorder entropy, leading to a temperature-dependent reduction of the hybridization gap \cite{Riseborough00, Riseborough92}. }
 
 \oli{The topological Kondo insulator must fulfill the vanishing hybridization at high symmetry points for all hybridization strengths, \ie for all temperatures.}
 \oliv{This can also be seen in the band structure with and without hybridization at $\Gamma$ and $X$, see  Fig. 1(d) of Ref. \cite{Takimoto11} and Fig. 3 of Ref. \cite{LuF13}.}  
    For smaller $V$ ---\ie at higher temperature--- the conduction band can assume a W-like profile with a local maximum at \X and higher binding energy valleys surrounding it.
    Upon lowering the temperature, a hybridization increase would transform it to become more U-shaped with an absolute minimum at \X  . These cases are sketched in Fig.~\ref{fig:schematic}(b).
    Such a change would give rise to a transfer of spectral weight to lower binding energy. Could imperfect \kperp resolution in the ARPES experiment integrate over a sufficiently large momentum range to give rise to the observed shifts? 
 
    In   Fig. \SuppFigRes\ we demonstrate that the answer is no. We approximate the ARPES intensity with a spectral function based on the conduction band dispersion shown in Fig.~\ref{fig:schematic}.
    Even if we assume a complete absence of \kperp resolution by integrating over the entire extent of the conduction band along $X$-$M$ [the \kperp direction for a (100) surface], we find that this cannot account for the observed shifts in Fig.~\ref{fig:bulktd}: While the transformation from a W- to U-shaped conduction band can cause a small upward shift, the experimentally observed result in Fig.~\ref{fig:bulktd} is 3 to 5 times larger.
    More importantly, the experimentally observed conduction band does not resemble the \kperp-integrated weakly hybridized profile. This is shown in Fig.~\ref{fig:cbshape}. The higher binding energy valleys around \X yield a plano-convex shape with the flat edge at the highest binding energy. In contrast, we exclusively observe a rounded underside experimentally for both $\kperp = 4\pi/a$ and $6\pi/a$.

We must therefore conclude that the observed shift of the three-dimensional photoemission feature at \Xbar results from a \emile{temperature}-induced shift of the conduction band dispersion at \X. 
\emile{This behavior directly rules out \smbsix as a topological Kondo insulator.}
\oliv{More precisely, one of the following applies: (i)  \smbsix is a hybridization insulator, but hybridization is nodeless and therefore not mixing opposite parity states to give a nontrivial topology. (ii) Alternatively, \smbsix becomes insulating by a non-rigid band shift that is unrelated to hybridization.
 In either case \smbsix is not a topological Kondo insulator.}

\subsection{Surface states}

\oliv{Numerous reports have claimed that the \emile{experimentally observed surface states are of topological nature.} In the following, we briefly discuss several open issues concerning these states, highlighting the lack of compelling evidence for \emile{this interpretation}.}

\subsubsection{Spin-polarized Photoemission}

\begin{figure}
    \centering
    \includegraphics[width=0.95\linewidth]{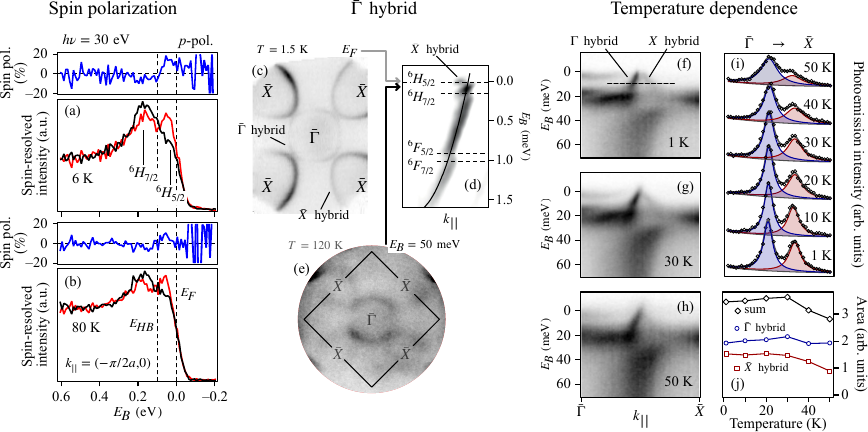}
    \caption{(left) Spin-polarized photoemission at 6 and 80 K. (center) The Sm $5d$ around \X at $\mathbf{k} = (0, 0, \pi)$ gives rise to a fourfold-symmetric surface hybrid near \EF analogous to the elliptical \Xbar surface hybrid. (right) Temperature dependence of the surface states reveals their persistence well beyond the closing of the bulk insulating gap.}
    \label{fig:2dsumm}
\end{figure}

In their review, Li \etal \cite{Li:2020jh} state that the observation of spin-momentum locking with spin-resolved ARPES is an essential argument in favor of the topological insulator scenario. We argue that there are significant doubts that this has been shown. In Fig.~\ref{fig:2dsumm} we show spin-resolved photoemission of the \Xbar state excited with   30~eV light (50~meV energy resolution). The light polarization is linear, which has been regarded as a key prerequisite for isolating the spin texture of the surface state \cite{Suga:2014vu,Li:2020jh}. The state attains a maximum spin polarization of 15\%. This is to the degree obtained by Xu \etal when using linearly polarized light ($\leq 15$\%, from Fig.~3 in Ref.~\cite{XuNC14}). The binding energy dependence of the spin polarization in Fig.~\ref{fig:2dsumm} shows a sign reversal at 100 meV, exactly as in the spin polarization due to the Sm $4f$ multiplet measured in Ref.~\cite{Suga:2014vu} \oliv{with circular light polarization}. \emile{This result suggests that the spin polarized emission we obtain with linearly polarized light is also due to the $f$ multiplet}.

Fig.~\ref{fig:2dsumm} further shows that this spin polarization persists from \oli{6 K} to 80 K. The small reduction at 80 K is due to the increased line width and resulting larger overlap and cancellation of oppositely spin-polarized channels. This result further shows that the origin lies in the atomic multiplet structure, not in the hybridized quasi-particle band structure. In the latter case, the spin polarization of a topological surface state must disappear with the closing of the hybridization gap as the temperature increases. Moreover, we find that the degree of spin polarization depends strongly on the photon energy and varies across the surface Brillouin zone.

Xu \etal, who measured at 20 K, argue that the spin polarization can be attributed to the dispersing surface states. Spin-resolved spectra (\ie energy distribution curves) are not shown in Xu \etal , but the spin polarization is shown for one higher binding energy. At 100 meV binding energy (Fig.~3k of Xu \etal ) the spin polarization vanishes, which led Xu \etal to conclude that the bulk states are not spin polarized \cite{XuNC14}. Our results in Fig.~\ref{fig:2dsumm}, however, show that the oppositely polarized contributions from the \sHfivetwo and \sHseventwo lines just cancel each other out near 100 meV binding energy. The spin-resolved ARPES spectra for the (111) surface by Ohtsubo \etal also fit our interpretation that emission from the $f$ multiplets is spin polarized \cite{Ohtsubo:2019hk}: In Fig.~4 of Ref.~\cite{Ohtsubo:2019hk} it can be seen that the spin polarization ---obtained with linearly polarized light of 26 eV--- does not tend to zero with increasing binding energy.

The attribution of the spin-polarized signal to the two-dimensional quasi-particle states near \EF is therefore not very convincing in either case. 

\subsubsection{`Umklapp' Intensity Reinterpreted}

Another unresolved issue is the interpretation of the largest electron pocket around \Gbar (labeled `\Gbar hybrid' in Fig.~\ref{fig:2dsumm}, referred to as $\beta '$ in Ref.~\cite{XuPRB13} and $\alpha '$ in Refs.~\cite{JiangNC13,Li:2020jh}). This feature resembles a mirror image of the elliptical Fermi surface contour at \Xbar . Ample evidence for a $(2 \times 1)$ surface structure~\cite{Yee:2013ve,Rossler:2014kn} led several groups to attribute this feature to a zone folding effect (back-folding of the \Xbar state, also referred to as umklapp peak)  \cite{JiangNC13,XuPRB13,Hlawenka:2018fj}.

Ohtsubo \etal (see Supplementary Note 10 of Ref.~\cite{Ohtsubo:2019hk}) and Li \etal \cite{Li:2020jh} challenge this interpretation with two convincing arguments: Firstly, the $(2 \times 1)$ reconstruction that is the basis for this assignment, is exclusive to Sm terminated surfaces, see Ref. \cite{Herrmann:2020ju}. The suggested umklapp intensity, on the other hand, is most clearly observed for boron terminated samples which appear as $1\times1$ in STM. Secondly, a more careful examination reveals that the feature's dispersion is not actually a mirror image of the \Xbar state: The Fermi surface contours and Fermi velocities differ \cite{Ohtsubo:2019hk,Li:2020jh}. 
Ohtsubo \etal and Li \etal interpret this feature  as the expected \Gbar centered topological surface state instead of umklapp \cite{Li:2020jh}. This notion is echoed in a layer-resolved theoretical study by Thunstr\"om and Held employing   density functional theory with dynamical mean-field theory \cite{Thunstrom:2021aa}.  

Here, we provide a third possibility: We will argue that it is the \Gbar analogue of the \Xbar contours.
In Ref.~\cite{Hlawenka:2018fj} we demonstrated that the elliptical \Xbar Fermi surface contour is due to a surface Sm $d$-$f$ hybrid in which the hybridization region is energetically shifted with respect to its bulk counterpart. Near \EF , where the hybrid has nearly complete $4f$ character, the spatial confinement and minimal bandwidth ensure that there is no mixing with its \threedim analogue. The surface hybrid therefore appears as a \twodim object near \EF .

If this explanation accounts for the surface states at \Xbar , we should also expect such a surface hybrid at \Gbar where the bulk $X$ points at $\mathbf{k} = (0,0, \pm \pi)$ are projected in the surface Brillouin zone. \oliv{(Note that we interpret in the same way the three surface states at the \Mbar\ points of (111) observed by Ohtsubo \etal \cite{Ohtsubo:2019hk}.)} We therefore propose that the $\alpha'$ feature, previously interpreted as umklapp, is in fact the expected \Gbar surface hybrid.

We prove this in Fig.~\ref{fig:2dsumm}.  \oli{Figure~\ref{fig:2dsumm}(c)} shows the Fermi contour, and \oli{Fig.~\ref{fig:2dsumm}(d)} shows the dispersion of the \Xbar state where the energies of the $f$ multiplets can be well distinguished by their hybridization. We see that the large contour around \Gbar does not exclusively appear at the Fermi energy where the Kondo gap opens but also at higher binding energies where hybridization with other multiplet lines occurs, see \oli{Fig.~\ref{fig:2dsumm}(e)}. This is exactly the behavior that is observed for the $d$-$f$ hybrid at \Xbar \cite{Hlawenka:2018fj}.  

We conclude that the bulk constant energy surfaces ---including the Fermi surface--- form \twodim $d$-$f$ hybrids that explain the Fermi level contours at \Xbar as well as at \Gbar. \oliv{The latter is crucial since there is an even number of \Xbar points in the surface Brillouin zone but the predicted strong topological insulator requires an odd number of topological surface states. }   

\subsubsection{Surface State Temperature Dependence}

Tunneling spectroscopy studies suggest the emergence of surface states concurrent with the opening of the bulk band gap \cite{Pirie:2019gr, Aishwarya22,Pirie:2023aa}. In Fig. 4 of Ref. \cite{Pirie:2019gr}, a Dirac cone is proposed \oli{based on quasiparticle interference}. This dispersion \oli{assigned to \Xbar} has been highlighted as a key feature of the topological Kondo insulator property of \smbsix in a recent review \cite{Checkelsky:2024aa} \oli{although the \Xbar\ point is not relevant for the topology of \smbsix and the crucial \Gbar Dirac cone is virtually absent.}

Pirie \etal also present temperature-dependent data that are not momentum-resolved. The signal attributed to the surface state decreases to half of its intensity between 2 K and 6-10 K. This is interpreted as the decay of the spectral weight of Dirac surface states as the Kondo insulator gap closes \cite{Pirie:2019gr}. Simultaneously, Pirie \etal claim excellent agreement between their signal and the size of Fermi surface contours observed in ARPES. They suggest that all other discrepancies with ARPES are due to the fact that only STM samples a single non-polar surface, while ARPES averages over a mixture of terminations \cite{Pirie:2019gr}.

In reality, we probe by ARPES boron $(1 \times 1)$ and samarium $(2 \times 1)$ terminations that are chemically pure on a macroscopic scale of hundreds of micrometers \cite{Hlawenka:2018fj,Herrmann:2020ju}.
Therefore, the data presented by Pirie \etal should be comparable to our ARPES results, including their temperature dependence.
\oli{Actually, Neupane et al. \cite{Neupane13} previously reported that in ARPES the signal of the surface states within the Kondo gap is suppressed before reaching 30 K, a claim that has not been contested yet.}

In Fig.~\ref{fig:2dsumm}\oli{(e-i)}, we present the temperature dependence of the surface features crossing \EF seen with ARPES: The Rashba split surface state centered at \Gbar and the two surface hybrids centered at \Gbar and \Xbar. The spectral weight shows minimal dependence on temperature in the range where the bulk gap is established. Notably, the ARPES features are clearly identified as surface-derived via photon-energy dependent ARPES. The observed temperature dependence of the surface states (\ie the surface hybrids) does not support a scenario in which they emerge upon the establishment of the bulk insulating phase.

\section{Discussion}

We find that the \threedim band structure of \smhb violates the rules of a topological Kondo insulator. \oliv{The non-rigid shift of the bulk $\Gamma$ and $X$ points indicates either a trivial Kondo insulator or an insulator due to a non-rigid band structure change that is \emph{not} hybridization and hence not Kondo like.} 
     
We now discuss the conclusions that can be drawn from our experimental work in relation to the theoretical predictions, as no theoretical study has predicted \smbsix to be trivial. We focus on the three most influential calculations and assess how far their predictions are from the trivial case.

First, the work by Takimoto  \etal \cite{Takimoto11}, which presented the first published surface band structure, predicted an inverted bulk band gap of only \oliv{2.4} meV, i.e., not far from the trivial case.

Second, the work by F. Lu  \etal \cite{LuF13} provided a Fermi surface much more similar to the ARPES experiment and offered a clear prediction for both surface and bulk band dispersion. However, by the same group and method, YbB$_6$ was also predicted to be topologically nontrivial \cite{Weng:2014bm}, a prediction that has since been disproven by ARPES experiments \cite{FrantzeskakisYbB6, Kang:2016bk}. YbB$_6$ is also not mixed valent.

Third, Alexandrov  \etal \cite{Alexandrov13} employed a large fermion-flavor number slave-boson mean-field theory, which had proven successful in heavy fermion superconductors \cite{Flint08}, to derive a topological phase diagram as a function of Sm valence. \oliv{They predict a topological insulator for Sm valences $>$2.56. Our measurements of the valence at low temperature in   Fig.~\ref{fig:bulktd}(d) place SmB$_6$ on the trivial side in the phase diagram. 
}

Topological insulators and semimetals have been extensively and successfully described in great detail using density functional theory, which has been crucial in overcoming the limitations of low-energy effective models \cite{Vergniory:2019aa}. In contrast, strong electron correlations remain a significant challenge for theoretical models. At the intersection of these two fields, it seems that the situation is more aligned with the challenges posed by electron correlation. For example, an established topological superconductor has yet to be identified.
It has been pointed out that there is no complete agreement on the minimal model for \smbsix and that, depending on the model, additional topological phases occur 
\cite{Baruselli15, Legner:2015hr}. Still, among the substantial literature, including most recent works \cite{Thunstrom:2021aa,LiuSchnyder23,Iraola24}, there has been no calculation suggesting that \smbsix is topologically trivial. Investigating the underlying reasons for this discrepancy could serve as a valuable first step in advancing our understanding.

\section{Methods}

Samples were grown by a crucible-free floating zone technique, cut, and oriented as described previously \cite{Hlawenka:2018fj}. 
Photoemission measurements were done using the 1$^3$-ARPES endstation at the UE112-PGM2b beamline at BESSY II. Samples are cleaved in ultrahigh vacuum at temperatures below 45 K. Spatial resolution is determined by the size of the
synchrotron beam with a full width at half maximum (FWHM) of $\approx250$ $\mu$m. The energy resolving power $E / \Delta E$ exceeds 10$^5$.

{\oli{The spin-resolved ARPES spectra were measured at the Spin-ARPES station at the U1252-PGM beamline of BESSY II. Photoelectrons were detected with a Scienta R4000 electron analyzer. For spin analysis, a Rice University Mott-type spin polarimeter was used, operated at 25 kV and detecting the in-plane component of the spin polarization. Resolutions of spin-resolved ARPES measurements were 0.75$^\circ$ (angular) and 50 meV (energy).}}

\oli{
Additional ARPES experiments were carried out at the NanoESCA beamline at Elettra, using a photon energy of 30 eV with variable horizontal, vertical, and elliptical polarization. The NanoESCA endstation hosts an electrostatic photoelectron emission microscope equipped with a double-pass hemispherical analyzer. The instrument, in reciprocal space mode operation (k-PEEM), is capable of detecting angle (momentum)-resolved photoemission intensities simultaneously in the full-emission hemisphere above the sample surface. An electron-optical column collects the photoemitted electrons, which are subsequently energy-filtered in the double-hemispherical configuration and finally projected onto a 2D detector.} 

\oli{XAS was performed at the High-Field Diffractometer station of the UE46-PGM1 undulator beam line of BESSY II \cite{Weschke18}. Spectra were taken around the Sm $M_{4,5}$ edge. To determine the temperature-dependent valence, Sm $M_5$ XAS spectra were fitted by a sum of theoretical spectra of Sm$^{2+}$ and Sm$^{3+}$ \oliv{in a way very similar to  the procedure described in Ref. \cite{Zabolotnyy18}}. The weighting coefficient was interpreted as the mean valence with a systematic uncertainty of $\pm0.05$. }


%

\end{document}